\documentclass[useAMS,usenatbib,usegraphicx,twocolumn]{mn2e}

\newcommand{\degree}{{\ensuremath{^\circ}}}

\def\lapp{\ \lower 3pt\hbox{${\buildrel < \over \sim}$}\ }
\def\gapp{\ \lower 3pt\hbox{${\buildrel > \over \sim}$}\ }

\newcommand{\be}{\begin{equation}}
\newcommand{\ee}{\end{equation}}
\newcommand{\bea}{\begin{eqnarray}}
\newcommand{\eea}{\end{eqnarray}}
\newcommand{\next}{\nonumber\\}
\newcommand{\rn}[1]{(\ref{#1})}

\title[Tidal evolution of GJ~436b]
{On the long-term tidal evolution of GJ~436b in the presence of a resonant companion}
\author[Rosemary A. Mardling]{Rosemary A. Mardling$^{1}$\thanks{E-mail:
mardling@sci.monash.edu.au}\\
$^{1}$School of Mathematical Sciences, Monash University, Victoria, 3800, Australia}
\begin{document}

\date{Accepted ... Received ...; in original form ...}

\maketitle

\begin{abstract}

In order to explain the significant orbital eccentricity of the short-period
transiting Neptune-mass planet GJ~436b
and at the same time satisfy various observational constraints and anomalies,
Ribas, Font-Ribera and Beaulieu
have proposed the existence of an eccentric low-mass companion planet at the position
of the outer 2:1 resonance.
The authors demonstrate the viability of their proposal
using point-mass three-body integrations,
arguing that as long as the system appears to be dynamically stable, the
short-term secular variations ought to dominate the long-term dissipative evolution.
Here we demonstrate that if one includes tidal dissipation, {\it both} orbits
circularize after a few times the circularization timescale
of the inner planet. 
We conclude that with or without a nearby 
companion planet, in or out of the 2:1 resonance, the $Q$-value of GJ~436b 
must be near the {\it upper} bound estimate for Neptune if the system is as
young as 1 Gyr, and an order of magnitude higher if the system is as old as 10 Gyr.
We show detail of passage through resonance and conclude that even out of
resonance, a companion planet should still be detectable through
transit timing variations.

\end{abstract}


\section{Introduction}

GJ~436b was discovered in a radial velocity survey by \citet{butler}, and has since
been observed in transit by \citet{gillon} and \cite{alonso} as the only transiting hot Neptune
to date.
Its system parameters
are listed in Table~\ref{orbit} together with those for the hypothetical
companion of Ribas, Font-Ribera \& Beaulieu (2008).
Here $m$ and $R$ are the mass and radius of the body \citep{deming},
$a$ is the semimajor axis, $e$ is the orbital eccentricity, $\omega$, $\Omega$ and $i$
are the argument of periastron, longitude of the ascending node and inclination respectively,
the latter two measured relative to the line of sight, $M$ is the mean anomaly, $Q$ is the $Q$-value,
$k$ is the Love number and $R_g$ is the radius of gyration. All angles are in degrees;
round brackets refer to the hypothetical orbital data used by \citet{ribas}, 
while square brackets refer to
initial data used in
the simulation presented here (see Section~\ref{two}).
\begin{table*}
\label{orbit}
\centering
 \begin{minipage}{140mm}
  \caption{Observed and hypothetical orbital and structural data for a GJ~436 two-planet system}
  \begin{tabular}{lccccccccccc}
  \hline
 & $m$ & $R$ & $a$ (AU) & $e$ & $\omega$ & $\Omega$ & $i$ &  $M$ & $Q$ & 
 $k$ & $R_g/R$\\
 \hline
 star & $0.452\, {\rm M}_\odot$ & $0.452\,{\rm R}_\odot$ &&&&&&& [$10^5$] & 
 [0.028]& [0.276]\\
 GJ~436b & $23.2\, {\rm M}_\oplus$ & $27,600\,{\rm km}$ &0.0287 & 
 0.15 & $343$ &  & $86.54$  &  &  &  & \\
 & &  &[0.0295] & 
 [0.2] &[$343$]  &[$90$] & [$86.54$]  & ($0$) & ? &[0.346] & [0.511]\\
 GJ~436c & ($4.7\, {\rm M}_\oplus$) & - & (0.045) & 
 (0.2) & ($265$) & [$30$] & ($96.54$) & ($107.6$)  & - & - & -\\
\hline
\end{tabular}
\end{minipage}
\end{table*}
Note that the mass and radius of GJ~436b are respectively 1.35 and 1.094 times those of Neptune.
As an extremely short-period system with a significant eccentricity, it has attracted 
considerable attention because it would appear that such a system should long ago
have circularized. However, there is one factor in particular which works against the intuition
that GJ~436b ought to be circularized, and that is that the planet's distance from the 
star in units of its {\it radius} is quite large at 160 compared to, for example, HD 209458b
at 76.
Since the circularization timescale depends on the fifth power of this ratio, it turns out to
be relatively long if one uses estimates for the $Q$-value of Neptune.
The latter has been estimated 
by \citet{banfield} to be in the range
$1.2\times 10^4<Q_N<3.3\times 10^5$, while \citet{tittemore} estimate the $Q$-value of
Uranus, whose mass is 0.85 times that of Neptune, to be less than $3.9\times 10^4$.
Using the expression
\be
\tau_{circ}=e_b/\dot e_b\simeq \frac{2}{42\pi} \left(\frac{Q_b}{k_b}\right)\left(\frac{m_b}{m_*}\right)
\left(\frac{a_b}{R_b}\right)^5 P_b
\label{taucirc}
\ee
for the circularization timescale of a synchronous system \citep{goldreich},\footnote{Note that
\citet{goldreich} use a modified $Q$-value, $Q'_b$, which absorbs
the Love number such that $Q_b'=3Q_b/2k_b$.}
where the subscript $b$ refers to quantities associated with GJ~436b, $m_*$
is the stellar mass and and $P_b=2.64\,{\rm d}$ is the orbital period,  one obtains the range
$5.3\times 10^7\,{\rm yr}<\tau_{circ}<1.5\times 10^9\,{\rm yr}$, where we have used the \citet{banfield}
estimates for Neptune as well as Jupiter's quadrupole Love number (0.34).
Being estimated to lie in the range 1-10 Gyr, the age of the system is quite uncertain \citep{torres}.
Thus in order for the non-zero eccentricity to be simply a result of a circularization time which is longer
than the age of the system, this
simple analysis suggests that $Q_b$ must be greater than $2.3\times 10^5 (\tau_{age}/{\rm Gyr})$,
where $\tau_{age}$ is the age of the system.

In this Letter we focus on the proposal of \citet{ribas}
that, given the circularization timescale is considerably less than the age of the system,
the eccentricity is sustained by the presence of a low-mass companion planet positioned
at the location of the outer 2:1 resonance, in an orbit which is inclined at around $10^\degree$
to that of GJ~436b. The authors claim to have found 
a strong peak at 5.2 days in a periodogram
analysis of the RV data, consistent with a body in the 2:1 resonance and with
a false alarm probability of 20\%. This is supported by a fit to the residuals of the
two-body fit to the inner orbit.
Moreover, the authors claim that harbouring a planet in the 2:1 resonance allowed them to produce the
observed eccentricity of GJ~436b with a planet whose mass is low enough not to have previously
been detected in the RV data.\footnote{Note that we find a maximum value of only 0.1 for $e_b$ over
a modulation cycle using their initial conditions, independent of initial phases and longitudes.}
Their scenario was further strengthened by the fact that it provides a natural
explanation for the apparent change in the inclination to the line of sight
of the orbit of GJ~436b, $i_b$, which would bring it into
transit between the time of the null result of \citet{butler} and the positive detection of \citet{gillon}.
The authors estimated that a rate of change of $i_b$
of around $0.1^\degree\,{\rm yr}^{-1}$ was needed
to be consistent with these observations (note that this generally includes a combination of nutation
{\it and} precession of the orbital plane).
However, \citet{alonso} have more recently reported an upper limit for the current rate of change of $i_b$ 
of $0.03\pm 0.05^o\,{\rm yr}^{-1}$. Since 
$0<di_b/dt<0.03^o\,{\rm yr}^{-1}$ for about 20\% of the precession cycle
(see Figure~\ref{gj2}(c)), this in itself does not rule out a Ribas et al-type system.

In the following section we briefly consider the tidal evolution of a single-planet system, putting a
more accurate lower bound on $Q_b$ for that case. In Section 3 we
study the proposed orbital solution of  \citet{ribas} using
a direct integration scheme for three bodies which includes perturbing accelerations due to tidal
dissipation and spin-orbit coupling in both the star and the innermost planet, 
and the post-Newtonian relativistic contribution to
the potential of the star \citep{mardlinglin}. 
We demonstrate that long-term evolution tends towards the doubly-circular state, independent
of the tidal circularization timescale of the outer planet, 
with the orbits inclined to each other by a fixed angle.
Section 4 discusses stability while Section 5 presents a conclusion.

\section{GJ~436b without a companion}\label{without}

\begin{figure}
\centering
\includegraphics[width=55mm]{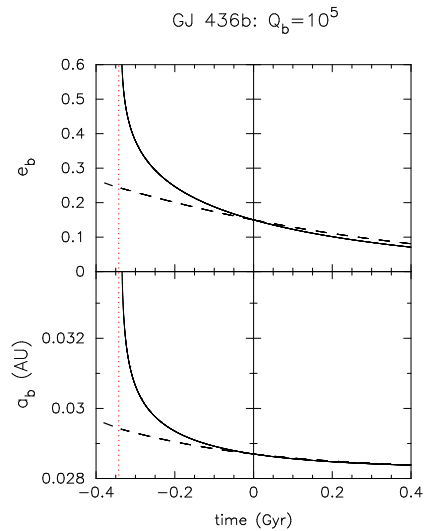}
\caption{Past and future evolution of GJ~436b without a companion.
The system cannot be older than $0.38(Q_b/10^5)\,{\rm Gyr}$.}
\label{plot2}
\end{figure}

The simplest explanation for the significant orbital eccentricity of GJ~436b is that
the circularization timescale is longer than the age of the system. Figure~\ref{plot2}
shows the results of integrating the secular equations \citep{mardlinglin} for a single synchronized
planet backwards
and forwards in time, with system parameters taken from Table~\ref{orbit} and $Q_b=10^5$
(solid curves).
The equations used are based on the analyses of \citet{hut}
and \citet{eggleton} 
in which a constant
time lag is assumed for the tidal bulge, and are correct for
any eccentricity (in as far as the constant time lag assumption is correct). 
If one assumes that no forces other than those included in the simulation were acting
in the past, and also that the $Q$-values of the planet and star as well as their
radii have not changed,
these results suggest that the system cannot be older than $0.38(Q_p/10^5)\,{\rm Gyr}$.
This is in contrast to the results obtained if one integrates equations which are correct to second-order
only in the eccentricity (dashed curves). The latter approximation is clearly adequate
for estimating the current circularization timescale, but is entirely inadequate for
understanding the past evolution. In particular, the second-order analysis does not allow
an upper bound estimate of the age of the system.
While alternative tidal models such as those which assume
constant lag angles for all tidal components \citep{goldreich},
or those which include the dynamical tide for high eccentricity \citep{mardlingtc}
may produce slightly different functional dependence on the eccentricity were the
analysis to be carried out, they are not likely to change the estimated age of the system by much.

Other forces are likely to have been operating early in the life of the system, including
those due to the rapid rotation of the star, larger stellar and planetary radii, the 
presence of a protoplanetary disk, coupling of the stellar and planetary magnetic fields \citep{laine},
and different $Q$-values.
Such factors will have determined the ``initial'' orbital state of the system, that is,
the state of the system when such influences became negligible, and it seems likely that
the eccentricity would not have been much larger than its present value if the single-planet scenario 
is correct.

The main conclusion one can draw from this simple analysis is that if GJ~436b does not have companions
capable of sustaining the observed eccentricity, the upper bound on its age in terms of $Q_b$
puts a {\it lower} bound on the $Q$-value of GJ~436b of $Q_b>3\times 10^5 (\tau_{age}/{\rm Gyr})$,
a value slightly higher than that gleaned from equation~\ref{taucirc}. In particular, it is 
approximately equal to the {\it upper} bound for the $Q$-value of Neptune if the system
is as young as 1 Gyr, and an order of magnitude larger if it is as old as 10 Gyr.

\section{Long-term tidal evolution of two-planet systems}

The second simplest 
explanation for a non-zero eccentricity in a system like GJ~436 is the presence of a companion
planet or star. In general the induced eccentricity of a stable system
will vary quasi-periodically, with periods of variation dominated by the rate of change of the
angle between the apsidal lines, the rate of change of the resonance angle(s)
if the system is in a resonance, and the rate of change of the {\it argument} of
periastron if the system is significantly non-coplanar \citep{murray}.
The amplitude of variation depends on the ratios of semimajor
axes and planet masses as well as the initial eccentricities and the angle
between the apsidal lines, and when relativistic effects are important, on the ratio of the inner
planet mass to the stellar mass. Formulae for the amplitude and period of variation of the eccentricities are
given in \citet{mardling-puff} for coplanar
non-resonant systems with moderate inner eccentricity ($e_b\lapp 0.2$). The
expression for the amplitude is also accurate for moderately non-coplanar
and/or resonant systems, while that for the modulation period
severely overestimates the true value for a system in or near the 2:1 resonance because for
such a close system, higher order and resonant terms should be included.

A mistake often made is that secular variations of the orbital elements persist for the lifetime
of the system, even when dissipative forces are significant \citep[eg.][]{maness}. 
In fact such variations are damped out after a few circularization timescales, 
with the system evolving to a
pseudo-equilibrium configuration which itself evolves on a generally longer timescale.
A secular theory for two-planet coplanar systems with dissipation
has recently been developed \citep{mardling-puff},
in which it is shown that the eccentricity of the innermost planet together with the angle between
the apsidal lines, $\eta$, evolve towards a fixed point in $(e_b,\eta)$ space on a timescale
of three times the circularization timescale of the inner planet, after which the system
evolves to the doubly circular state on timescale given by equation~(60) in \citet{mardling-puff}.
{\it Note that this timescale is independent of the $Q$-value of the outmost planet}.
When resonant terms are important and/or the system is moderately inclined, the behaviour
is only slightly modified (as long as the system is stable). 
The fixed point or equilibrium eccentricity, {\it a quantity which is independent
of the initial values of $e_b$, phases and longitudes}, is given by
\be
e_b^{(eq)}=\frac{(5/4)(a_b/a_c)\,e_c\,\varepsilon_c^{-2}}
{\left|1-\sqrt{a_b/a_c}(m_b/m_c)\varepsilon_c^{-1}+\gamma\varepsilon_c^3\right|},
\label{equilGR}
\ee
where $\varepsilon_c=\sqrt{1-e_c^2}$
and $\gamma=4(n_b a_b/c)^2(m_*/m_c)(a_c/a_b)^3$, with $n_b$ the mean motion
of the inner planet and $c$ the speed of light.\footnote{The quantity $\gamma$ is the 
ratio of the relativistic to the three-body contribution to the rate of apsidal advance of 
the innermost planet.}
Note, however, that \rn{equilGR} was derived assuming that the average
value of the outer eccentricity doesn't vary much on 
the circularization timescale. When it does (as happens for the hypothetical GJ~436 system
because the outer mass is so low),
the estimate \rn{equilGR} tends to {\it overestimate} the equilibrium eccentricity. Thus \rn{equilGR}
can be regarded as an upper bound for $e_b^{(eq)}$.

\subsection{GJ~436b with a resonant companion}\label{two}
\begin{figure}
\centering
\includegraphics[width=55mm]{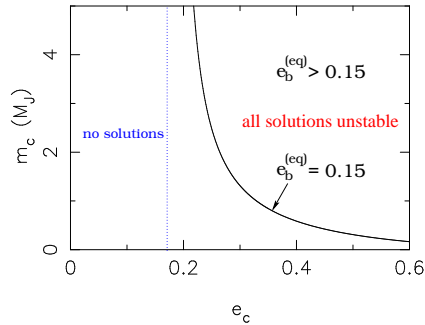}
\caption{Possible values for $m_c$ and $e_c$ corresponding to $e_b^{(eq)}=0.15$
with $a_c=0.045\,{\rm AU}$. No solutions exist for $e_c<0.18$, while essentially {\it all} of those with
$e_c>0.18$ are unstable.}
\label{equil}
\end{figure}

If the circularization timescale of GJ~436b is less than a third of the age of the
system, it will already have reached and evolved past the equilibrium eccentricity.
Thus we begin by calculating the range of companion masses and eccentricities
capable of producing an equilibrium eccentricity of 0.15, that is, the observed eccentricity of GJ~436b.
As in the previous Section, this will allow us to put a lower bound on its $Q$-value.
Using this value in \rn{equilGR} as well as a period ratio of 2,
we can rearrange the equation to write $m_c$ as a function of $e_c$. The result is plotted in Figure~\ref{equil}
which shows that no solutions exist for $e_c<0.18$. Solutions corresponding to
$e_c\gapp 0.18$ are actually unstable (see next section)
so that in fact {\it no} systems exist for which the equilibrium eccentricity
is equal to the observed eccentricity. 
(In fact, the equilibrium eccentricity corresponding to the hypothetical companion of \citet{ribas}
(with $e_c=0.2$) is 0.06.)

One can conclude from this that if GJ~436b {\it does} have a single
nearby low-mass companion in or near the 2:1 resonance, 
the system cannot yet have evolved to its pseudo-equilibrium
state. Its circularization timescale must therefore be longer than one third of the age of the system,
thereby providing the weaker constraint on the $Q$-value of GJ~436b than in the single-planet case that 
$Q_b>7.7\times 10^4(t_{age}/{\rm Gyr})$. 
In fact, we can do better than this, and argue that after only one circularization timescale the system
will be in apsidal libration with zero minimum $e_b$ and a maximum which depends on $e_c$
\citep{mardling-puff}.
From stability considerations $e_c$ can't be much more than around 0.3
(given it needs to be able to move safely through the resonance as it tidally evolves; see
Figure~\ref{stability}).
Since this corresponds to a {\it maximum} value of $e_b$ in a libration cycle of 0.12, we must have
that $\tau_{circ}>\tau_{age}$, putting the same lower bound on $Q_b$ as in the single-planet case.

We finish this section by demonstrating the behaviour discussed above.
Using the data in Table~\ref{orbit},
we performed a direct integration using the code described in the Introduction \citep{mardlinglin}.
The star's $Q$-value is an estimate, while the Love numbers $k$ (twice the apsidal motion
constant) and
radii of gyration $R_g$ for the star and planet correspond to $n=3$ and $n=1$ polytropes 
respectively \citep{sterne}, the latter often taken to
approximate the structure of Jupiter. 
The spin period of the star was taken to be 20 days while the spin of the planet was
taken to be synchronous with the orbital motion. Both were taken to be
aligned with the orbit normal.

\begin{figure}
\centering
\includegraphics[width=65mm]{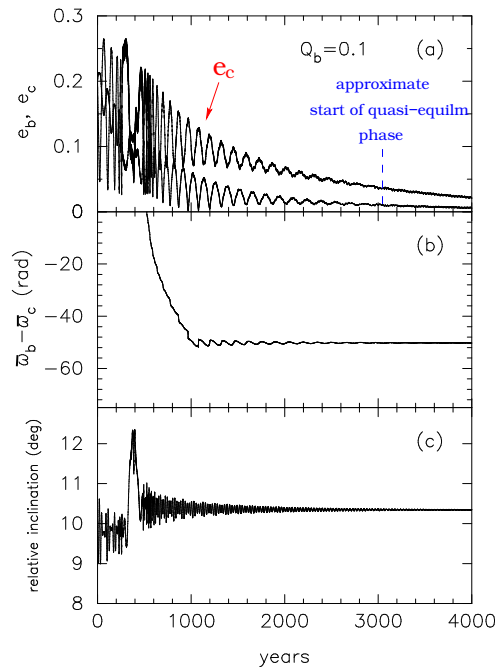}
\caption{Approach to the doubly circular state of a resonant two-planet GJ~436 system.
See text for discussion.}
\label{gj1}
\end{figure}

The secular analysis in \citet{mardling-puff} demonstrates that varying the $Q$-value of
the inner-most planet merely changes the timescale on which the system evolves towards
the final state while not affecting the local secular oscillation period.
Since tidal dissipation in the planet
dominates the tidal evolution while the inner orbit's eccentricity is non-zero, we took
$Q_b=0.1$ in order to show the detail of various stages of evolution, and confirmed that
the evolutionary timescale scales linearly with $Q_b$ for $Q_b=1$ and 10.
We also confirmed that the system remains stable for $2\times10^5\,{\rm yr}$ with $Q_b=10^6$,
consistent with the results of Section~\ref{stable}.
Figure~\ref{gj1}
shows 4000 years of evolution, equivalent to $4\,(Q_b/10^5)\,{\rm Gyr}$ for general $Q_b$,
except that $Q_b/0.1$ as many secular oscillations will have occurred in that time.
It passes through its present (hypothetical) configuration after $1600 (Q_b/0.1)\,{\rm yr}$.
Panel (a) shows the evolution of $e_b$ (bottom curve) and $e_c$, with the 
system entering resonance at around $200(Q_b/0.1)\,{\rm yr}$, and entering apsidal libration (panel (b))
when $e_b$ (temporarily) hits zero at $1000(Q_b/0.1)\,{\rm yr}$. The latter occurs on a
timescale of $1\,\tau_{circ}$ after which the system evolves towards the quasi-equilibrium phase on a timescale
of $2\,\tau_{circ}$.
The subsequent approach of both eccentricities to zero 
is well approximated by the secular theory and occurs on 
a timescale of around $6\,\tau_{circ}$ in this case \citep[equation~(60)][]{mardling-puff}.
Panel (c) shows the approach to a fixed value of around $10.3^\degree$ of the relative inclination.

\begin{figure}
\centering
\includegraphics[width=65mm]{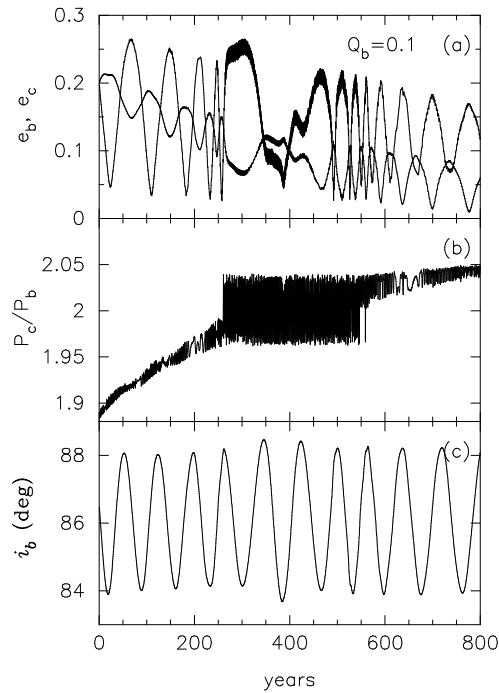}
\caption{Detail of the resonant phase of Figure~\ref{gj1}.
Note that the period and amplitude of these variations is independent of $Q_b$.
See text for discussion.}
\label{gj2}
\end{figure}

Figure~\ref{gj2}
shows detail of the passage through resonance. While the amplitude of variation of the eccentricity
is only slightly enhanced (so that estimates provided by the secular theory are reasonably accurate),
the amplitude of variation of the orbital period ratio $P_c/P_b$
suddenly increases as the system crosses the separatrix and
enters the 2:1 resonance, with a width of around 0.035.
The corresponding variation of the orbital period of GJ~436b is around 15 mins, a significant
increase on the variation outside the resonance, whose effect on the transit timing would be
easily measured  (note that the libration period for this system is around 200 days). 
For coplanar systems in the 2:1 resonance with $m_b$, $m_c\ll m_*$,
this variation is given by (Mardling, in preparation)
\bea
\delta P_b/P_b&=&2\left[1+(m_b/m_c)\sigma^{2/3}\right]^{-1}
\delta\sigma/\sigma\next
&\simeq& 2^{-2/3}(m_c/m_b)\,\delta\sigma,
\label{Pi}
\eea
where $\sigma=P_o/P_i$,
$\delta\sigma=\sigma-2$, that is, the ``distance'' from resonance,
and the approximation holds for $m_c\ll m_b$.\footnote{See 
also \citet{holman} and \citet{agol} but note that our expression is independent of the
eccentricities, and is an approximation to an expression valid for any masses.}
However, the GJ~436 system passes through the resonance in around 
$300(Q_b/0.1)\,{\rm yr}=0.3(Q_b/10^5)\,{\rm Gyr}$ so unless $Q_b$ is at least
equal to the upper estimate of 
Neptune's $Q$-value, we would not expect to see the system in resonance now
(given that it was deposited into the resonance around the time of formation). \citet{alonso}
find no obvious departure from linear ephemeris and conclude that the proposed resonant
solution of \citet{ribas} is unlikely. Once the system leaves the resonance, the
amplitude of variation of $P_b$ is given by \rn{Pi} with $\delta\sigma$ replaced by the quantity
$\delta\sigma-\sqrt{\delta\sigma^2-\Delta\sigma^2}$, where $\Delta\sigma$ is the width of
the resonance (Mardling, in preparation). 
$\delta P_b$ reduces quickly as the system crosses the resonance,
however, even when the system becomes doubly circularized $\delta P_b$ is still significant
at around one minute. Thus we conclude that even out of resonance, the companion
planet proposed by \citet{ribas} would be detectable through transit timing variations given 
current accuracies \citep{alonso}.

Panel (c) shows the variation of the inclination of GJ~436b to the line of sight, $i_b$. The resonance
has very little effect on the period and amplitude of variation, with the average rate of change
of $i_b$ equal to approximately $0.06^\degree\,{\rm yr}^{-1}$. This compares to the observational
upper bound of 
$0.03\pm 0.05^\degree\,{\rm yr}^{-1}$ by \citet{alonso}. 

\subsection{Stability}\label{stable}

\begin{figure}
\centering
\includegraphics[width=60mm]{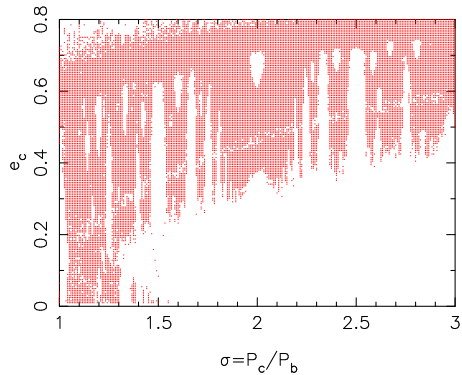}
\caption{A stability map of the region surrounding the 2:1 resonance for a range
of values of $e_c$.}
\label{stability}
\end{figure}

Figure~\ref{stability}
shows a stability map for a system composed of GJ~436b and a companion planet
with a mass of $4.7M_\oplus$, aligned periastra, a mutual inclination of $10^\degree$, and
with initial companion eccentricity $e_c$ 
and period $P_c=\sigma P_b$ indicated by the position in the plot. 
Three-body integrations were performed, with stability being determined
using the procedure 
described in \citet{cambody}.
Initial conditions corresonding
to unstable systems are indicated
in the figure by red dots; stable systems are left blank. 
A clearly defined boundary is evident, indicating that a system with a sufficiently low
initial value of $e_c$ would be free to tidally evolve through the 2:1 resonance without
the danger of instability.
Other resonances are also evident, for example, the 3:2 and the 5:2. 

\section{Conclusion}

The main conclusion from this study is that with or without a single nearby 
companion planet, in or out of the 2:1 resonance, the $Q$-value of GJ~436b must
be greater than the upper bound estimate for Neptune if the age of the system is around 1 Gyr,
and up to an order of magnitude greater for an age of up to 10 Gyr.
Passage through resonance of a Ribas et al.\ two-planet system
occurs on a timescale of $0.3(Q_b/10^5)\,{\rm Gyr}$, remaining stable throughout and beyond.
However, even it
were now no longer in resonance, the companion planet should still be detectable through
transit timing variations.

\label{lastpage}

\end{document}